\documentclass[a4paper,fleqn,usenatbib,useAMS]{mnras}

\usepackage{graphicx}	% Including figure files
\usepackage{amsmath}	% Advanced maths commands
\usepackage{amssymb}	% Extra maths symbols
\usepackage{multicol}        % Multi-column entries in tables
\usepackage{bm}		% Bold maths symbols, including upright Greek
\usepackage{pdflscape}	% Landscape pages
\usepackage{tabularx} 

\newcommand{\simgt}{\hbox{\,\rlap{\raise 0.425ex\hbox{$>$}}\lower 0.65ex\hbox{$\sim$}\,}}
\newcommand{\simlt}{\hbox{\,\rlap{\raise 0.425ex\hbox{$<$}}\lower 0.65ex\hbox{$\sim$}\,}}
\newcommand{\grale}{{\tt Grale}}
\newcommand{\lenstool}{\textsc{Lenstool}}
\newcommand{\arcsecond}{$^{\prime\prime}$}

\usepackage[T1]{fontenc}
\usepackage{ae,aecompl}

\title{Testing light-traces-mass in Hubble Frontier Fields Cluster MACS-J0416.1-2403}

\author[K. Sebesta et al.]{Kevin Sebesta$^{1}$\thanks{Contact e-mail: \href{mailto:sebesta@physics.umn.edu}{sebesta@physics.umn.edu}}
Liliya L. R. Williams,$^{1}$
Irshad Mohammed,$^{2,3,4}$
Prasenjit Saha$^{2,3}$ 
\newauthor ~and~ Jori Liesenborgs$^{5}$
\\
$^{1}$School of Physics \& Astronomy, University of Minnesota, 116 Church Street SE, Minneapolis, MN 55455, USA\\
$^{2}$Physik-Institut, University of Zurich, Winterthurerstrasse 190, 8057 Zurich, Switzerland\\
$^{3}$Institute for Computational Science, University of Zurich, Winterthurerstrasse 190, 8057 Zurich, Switzerland\\
$^{4}$Theoretical Astrophysics Group, Fermi National Accelerator Laboratory, Batavia, IL 60510, USA\\
$^{5}$Expertisecentrum voor Digitale Media, Universiteit Hasselt, Wetenschapspark 2, B-3590, Diepenbeek, Belgium
}

\pubyear{2015}

\begin{document}
\label{firstpage}
\pagerange{\pageref{firstpage}--\pageref{lastpage}}
\maketitle

\begin{abstract}
We reconstruct the projected mass distribution of a massive merging Hubble Frontier Fields cluster MACSJ0416 using the genetic algorithm based free-form technique called Grale. The reconstructions are constrained by 149 lensed images identified by Jauzac et al. using HFF data. No information about cluster galaxies or light is used, which makes our reconstruction unique in this regard. Using visual inspection of the maps, as well as galaxy-mass correlation functions we conclude that overall light does follow mass. Furthermore, the fact that brighter galaxies are more strongly clustered with mass is an important confirmation of the standard biasing scenario in galaxy clusters. On the smallest scales, approximately less than a few arcseconds, the resolution afforded by 149 images is still not sufficient to confirm or rule out galaxy-mass offsets of the kind observed in ACO 3827. We also compare the mass maps of MACSJ0416 obtained by three different groups: Grale, and two parametric Lenstool reconstructions from the CATS and Sharon/Johnson teams. Overall, the three agree well; one interesting discrepancy between Grale and Lenstool galaxy-mass correlation functions occurs on scales of tens of kpc and may suggest that cluster galaxies are more biased tracers of mass than parametric methods generally assume.
\end{abstract}


%%%%%%%%%%%%%%%%%%%%%%%%%%%%%%%%%%%%%%%%%%%%%%%%%%

%%%%%%%%%%%%%%%%% BODY OF PAPER %%%%%%%%%%%%%%%%%%

\section{Introduction}
The phenomenon of bending of light due to the intervening matter, referred to as gravitational lensing, provides an opportunity to use massive clusters of galaxies as natural telescopes \citep{fm94,kn11}, and hence presents a unique tool to examine far away galaxies in detail. According to the standard $\Lambda$CDM model of cosmology, the first structures in the Universe form due to the gravitational instability of the dark matter and baryons which collapse under their own gravity \citep{alb61,pee70}. Later, massive galaxies and galaxy clusters formed by a series of mergers of the smaller systems \citep{pee83,blu84}. In this hierarchical structure formation scenario there exists a large population of low and intermediate mass galaxies at higher redshifts. If these happen to sit behind a massive galaxy cluster, their light will be bent, and their images distorted and magnified according to Einstein's Theory of General Relativity. This magnification, which can be significant, is the motivation behind the Hubble Frontier Fields \citep{bul12} program (hereafter HFF). The Space Telescope Science Institute selected six galaxy clusters, and associated parallel fields to gain a magnified view of the high redshift galaxies behind them, and to investigate complex galaxy clusters doing the lensing.

To use clusters as telescopes one must have a detailed and accurate characterization of their optics, or maps of the mass distribution. The present work is concerned with the mass distribution of one of the HFF clusters, MACS-J0416.1-2403 (hereafter MACSJ0416). It is a massive gravitational lens at a redshift of 0.396, and right ascension 04:16:09, and declination -24:03:58, \citep{man12}. The cluster shows features of a recent major merger or pre-merger \citep{jau15,ogr15}, like double-peaked X-ray surface brightness \citep{man12}, elongation, and many sub-structures \citep{zit13,gri15}. It has been studied in detail for its mass distribution \citep{zit13,jau14,jau15,gri15,die15}, magnification maps \citep{joh14,ric14}, and mass power spectrum \citep{moh16} using different inversion methods.

%If the lensing system's sky-projected surface mass density is larger than a critical value, the regime of strong gravitational lensing applies, and multiple images of each background galaxy (or the source) will be present. Therefore, it is possible to reconstruct the mass of the lens providing the multiple images of the background sources, this phenomenon is also referred to as lens reconstruction or inversion.

The goal of the lens reconstruction techniques is to build mass models of the lens constrained by the gravitational lensing data, which, for strong lensing consists of identifications, positions and redshifts of multiple images of background sources. There are a few different lens inversion methods in existence, based on one of the two different approaches to lens reconstruction --  parametric and non-parametric, or free-form. Some methods are hybrid \citep{die15}, and incorporate features of both approaches. Parametric methods assume an underlying functional form of the mass distribution of the lens, and the lensing data is used to constrain the parameters of the functional form. The number of parameters is usually small, and statistical inference can be made using regular sampling methods, like Markov-Chain Monte-Carlo (MCMC). Though in principle these physical models can take any form, in practice parametric models assign mass to cluster galaxies, assuming light-traces-mass, and add a few additional mass components to represent the cluster dark matter. One widely used example of such techniques is \lenstool~ \citep{jul07}. On the other hand, non-parametric models (or free-form) solve for the lensing mass by using the images data alone, with no reference to cluster galaxies. Some free-form methods match the number of free parameters to the number of lensing observables, while others work with a parameter space whose dimensionality greatly exceeds the number of constraints. An example of the latter type of method is \grale~\citep{lie06,lie07,lie08,lie09} which uses no light information from the lens, and requires only the position and redshifts of the multiple images of the background sources. 
%The total mass map is the superposition of building blocks, for example Plummer lenses, and \grale uses a genetic algorithm to optimise for their weights. 
In this paper we use \grale~to construct mass models of MACSJ0416.

Our paper has two goals: (i) to study the mass distribution of the cluster with respect to the light, and (ii) to compare results of two very different methodologies: free-form \grale~with $n_{param}\gg n_{constr}$, and light-traces-mass \lenstool, with $n_{param}\!\simlt\!n_{constr}$. To accomplish the first goal in an unbiased way one needs a method that does not include cluster galaxies as input; \grale~satisfies that criterion. One motivation for this is the recent discovery of an offset between a galaxy deep inside the gravitational potential of a massive cluster, ACO 3827 and its dark matter halo \citep{ws11,moh14,mas15}. Because our analysis is confined to the central portions of the cluster we do not expect to find mass-light discrepancies of the kind found by \cite{mer11} and \cite{jau15}. Our second goal has been addressed by some recent papers \citep{zit10,coe12}, but usually these compare circularly averaged density profiles, which is probably not the most suitable statistic in this case as MACSJ0416 underwent a recent merger. We use a more detailed metric, namely mass-galaxy correlation function.

In Section~\ref{sec:lit} we give a brief summary of the existing mass reconstructions of this cluster. In Section \ref{sec:grale} we discuss our lens reconstruction method. Section \ref{sec:results} presents our results, and in particular discusses light and mass offsets and galaxy-mass correlation function obtained with \grale, and two inversions with \lenstool. In Section \ref{sec:con} we summarize our findings. We use flat $\Lambda$CDM cosmology, with $\Omega_m=0.27$, and $h=0.71$.

\section{Existing Mass Models of MACSJ0416}
\label{sec:lit}

Current literature on mass reconstructions of MACSJ0416 have used free-form, hybrid and parametric methods to calculate the associated mass distribution. \cite{zit13} used two different parametric methods to determine the mass distribution of MACSJ0416. Their first method assigned a PIEMD (pseudo-isothermal elliptical mass distribution) mass profile to every galaxy and smoothed the total resulting mass map with an elliptical Gaussian to obtain a cluster-wide dark matter distribution. The latter was added to the mass due to individual galaxies. The second method used the same PIEMD model for galaxies, and two elliptical NFW halos to describe the dark matter. The authors ran an MCMC algorithm to obtain the solution for both methods. The large number of multiple images in this cluster was attributed to the extreme elongation of MACSJ0416. With their analysis the authors discovered 70 new multiple images from 23 sources.

\cite{joh14} modeled all 6 HFF clusters using pre-HFF data, and analyzed the resulting mass and magnification properties of the reconstructions. They followed the parametric approach by utilizing \lenstool~to find the best solution for the mass distribution in the clusters. They assigned a PIEMD to each cluster member, and to two cluster-scale components. The best solution was found through an iterative process of running an MCMC algorithm at each step. The final model was computed under image plane optimization, meaning the differences in the observed and model predicted positions were optimized in the lens plane. The authors report a smaller mass inside the $z = 2$ critical curve than \cite{zit13} because of different redshifts used in each model.

\cite{ric14} adopted a strong- and weak-lensing approach to find the mass distribution of MACSJ0416. They used \lenstool~with dual pseudo-isothermal elliptical (dPIE) mass profiles for each cluster member, and two cluster-wide mass clumps placed in the regions of high galaxy number density plus a third halo for a lower-redshift galaxy not belonging to the cluster. 

\cite{jau14} used \lenstool, which assigns a dPIE mass profile to cluster galaxies, and cluster wide components.  Based on the HFF data the authors identify 51 new multiply imaged systems, for a total of 68, comprising red 194 individual lensed images. Their best model produces lens plane rms of $0.68$\arcsecond~for 149 images of 57 most securely identified sources.

\cite{gri15} used a parametric method called GLEE. They considered a range of ways of parametrizing the mass distribution in the cluster, and found that the best fit was obtained with a model that has two cored elliptical pseudo-isothermal mass distributions to represent the dark matter and 175 galaxy-size dual pseudo-isothermal mass distributions with the mass-to-light scaling with luminosity.  Their model reproduces the positions of 10 image systems, totalling 30 spectroscopically confirmed lensed images very well, with lens plane rms of 0.3\arcsecond. Their results are largely in agreement with previous reconstructions of the shape of the mass distribution. A comparison with simulated galaxy clusters with total masses similar to that of MACSJ0416 shows that the former contain considerably less mass in subhalos in their cores relative to MACSJ0416.
 
\cite{die15} used a hybrid approach to reconstruct the mass distribution of MACSJ0416. Three separate mass models were made for the galaxy component of the lens, two following the light traces mass assumption and the third linking every galaxy to a circularly symmetric NFW distribution. The rest of the mass distribution was modeled by a free-form method.  These mass models showed a bimodal mass distribution, similar to the X-ray emission distribution, except for small
offsets in the two peaks. Collisional effects such as dynamical friction are believed to be the reason behind the offsets between X-ray and dark matter distributions. In addition, a flat mass profile was found on medium distance scales surrounding the two peaks of the mass distribution, likely because of tidal forces, projection effects, or possibly self interacting dark matter. The authors find that overall light traces mass in MACSJ0416.

\begin{table}
  \caption{Previous MACS0416 mass models with type of data used and number of multiple images.}
  \label{tab:mmodels} 
  \begin{tabularx}{\columnwidth}
    { l | l | X }
    \hline
    Author & Type of Data & Num. of Multiple Images \\ \hline
    
    \cite{zit13} LTM-Gauss & pre-HFF & 34 \\
    \cite{zit13} NFW & pre-HFF & 34 \\
    \cite{joh14} & pre-HFF & 34  \\
    \cite{ric14} & pre-HFF & 47 \\
    \cite{jau14} & HFF & 149 \\
    \cite{gri15} & CLASH & 30  \\
    \cite{die15} & CLASH & 96 \\ \hline
    
  \end{tabularx}
\end{table}

Published mass reconstructions of MACSJ0416 are summarized in Table \ref{tab:mmodels}. Besides the \cite{zit13} LTM-Gauss method, every previous parametric method used two cluster-scale haloes. All previous reconstructions made use of the galaxy light in MACSJ0416 in some form when calculating the mass distribution. We go about the reconstruction process differently, and use only the positions of the lensed images as input. Thus, light is not assumed to trace mass in our reconstruction processes, and our results represent a completely independent way of modelling MACSJ0416 and testing if light follows mass in clusters.

\section{{\Large {\grale}}: Free-form Lens Reconstruction Method}
\label{sec:grale}

\grale~is a free-form technique of lens reconstruction that uses a genetic algorithm to calculate the mass distribution of a lens. Only the images' identifications, locations and redshifts are used as inputs for \grale. The images' positions can be modeled either as point sources or as extended regions. For this paper we chose to use point sources to model the images. To start, a grid is set up with a uniform grid of cells. Within each cell a projected Plummer mass distribution is placed, fixing the width of each Plummer sphere. Then the genetic algorithm searches the parameter space of the Plummer amplitudes and determines the best solution to the lens equation and the corresponding mass distribution. A new grid is then formed from the original grid by increasing the number of cells in areas of high mass density. The genetic algorithm then calculates the mass distribution again. This process of increasing the density grid cells is repeated several times. Through each iteration, reproduction and mutation are used to breed new solutions. Thus, \grale~obtains the final mass distribution \citep{lie06}.

\grale~uses fitness values to pick the best solution for the mass distribution. The fitness value of a genetic algorithm evaluates the degree to which the mass map reproduces the lensing data, and hence if a particular solution is better or worse than others. In general \grale's fitness value can be based on one, two, or more fitness criteria; for this paper we chose to use two. One of the fitness value components depends upon image positions and the other uses the absence of images. For a given mass distribution, the lens equation can be used to project the images from the same source back to the source plane. \grale~uses the size of the region containing all back-projected images as a length scale. Thus an absolute scale is not used, because this would favor solutions that place the images into smaller source plane regions, and result in overfocussing. The first component of \grale's fitness value is calculated by measuring the distance between all back-projected images of the same source, using the aforementioned length scale. The second component is based on spurious extra images. In complex lensing systems, additional images could exist that trace back to sources. If the modeler is sure that these extra images do not exist, then the null space of \grale~is defined to be the area of no images in the image plane. The null space is divided in triangles and each triangle is back-projected to the source plane. The second component of \grale's fitness is the number of all the triangles that back-projects to the region containing all back-projected images. This paper takes the product of the two fitness components to find a single value to compare the different iterations of a given run.

We generated 30 \grale~reconstructions, each one using nine successive lensing grid refinements. We used 149 secure images as outlined in \cite{jau14} and updated redshifts from \cite{gri15}. For each grid the number of Plummer distributions was chosen at random from a selected range. The first lensing grid's range was 300-400 and the last, ninth lensing grid's range was 1700-1800. Each subsequent grid had a greater density of Plummer spheres. In addition, the center coordinates and size of each grid was allowed to fluctuate by 5\arcsecond~and 10\arcsecond~respectively to eliminate the imprint of a fixed grid on the solution. Finally, for each run the best mass map was selected depending upon the fitness value described above. Each reconstruction's results are somewhat different because the random seed used by \grale~to initialize a run enables it to explore different regions of the large dimensional model parameter space.  To reduce the random variations and to enhance the common features in mass maps of different runs, the runs---30 in our case---are averaged to produce one final mass map, which is also a solution, as discussed in \cite{moh14}.

\section{Results}
 \label{sec:results}

 The average map of the 30 individual \grale~reconstructions is shown in Fig.~\ref{fig:avgmap}; this map is the basis of our analysis. The blue lines are the mass density contours and the red circles indicate images. There are 20 contour lines linearly spaced from 0.03827 g/cm$^2$ up to 0.76531 g/cm$^2$. Using the 30 individual maps we can calculate the fractional uncertainty defined at every location in the lens plane as the ratio of the root-mean-square deviation between the maps' $\kappa$ values and the average $\kappa$ of all the maps. ($\kappa$ is the projected surface mass density in units of critical density for lensing). The fractional uncertainty highlights areas in the mass distribution that contain small and large uncertainty. As seen in Fig.~\ref{fig:fracmap}, there are some small areas within the central parts of the galaxy cluster that contain large fractional uncertainty, 30-40\% (where contour lines are green and the density of lines is greater), however most of the galaxy cluster is minimal in fractional uncertainty, usually below 15\% and in some regions below 5\% (where contours are thin and the density of lines is lower), because \grale~is very well constrained by the images in this region. Large fractional uncertainty values can be seen outside the central elongated region of the cluster where \grale~does not have many constraints.

\grale, or any other lens reconstruction method that does not explicitly place mass at the location of visible galaxies as part of its input is the right tool for investigating how well mass follows light on the scale of galaxies in the cluster. No other published lensing inversion of MACSJ0416 is form-free in this sense. Here, we investigate this using two approaches, described in the next subsections.

\subsection{Mass Contours and Local Mass Peak Offsets}
\label{sec:peaks}

\begin{figure*}
  %\centering
  \includegraphics[width=\textwidth]{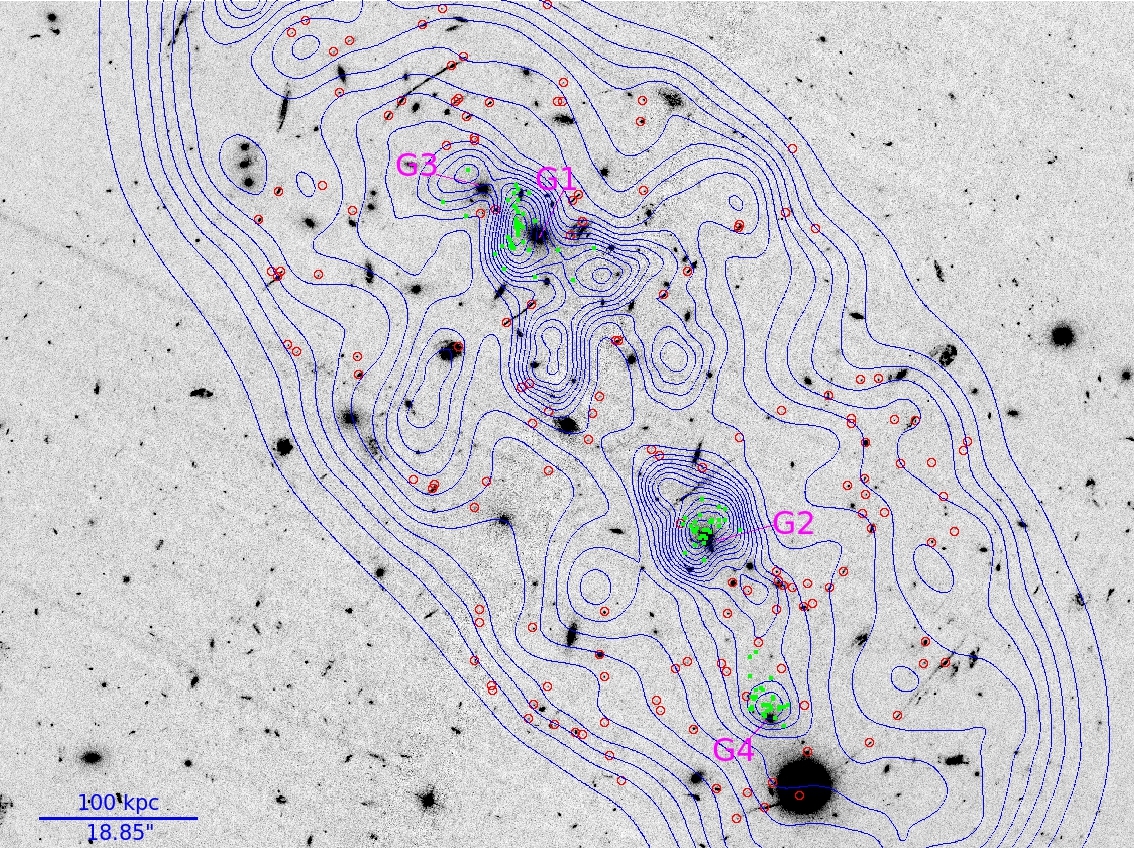}
  \caption{Mass contours of averaged mass map of MACSJ0416 red overlaid on a HST F435W image. The blue lines represent the mass density contours and the red circles are images. There are 20 mass contour levels linearly spaced from 0.03827 g/cm$^2$ up to 0.76531 g/cm$^2$. red The green dots represent local mass peaks around the 4 central ridge galaxies, in 30 individual mass reconstructions. These galaxies are labelled G1-G4, with G1 appearing the brightest.\label{fig:avgmap}}
\end{figure*}

\begin{figure*}
  \includegraphics[width=\textwidth]{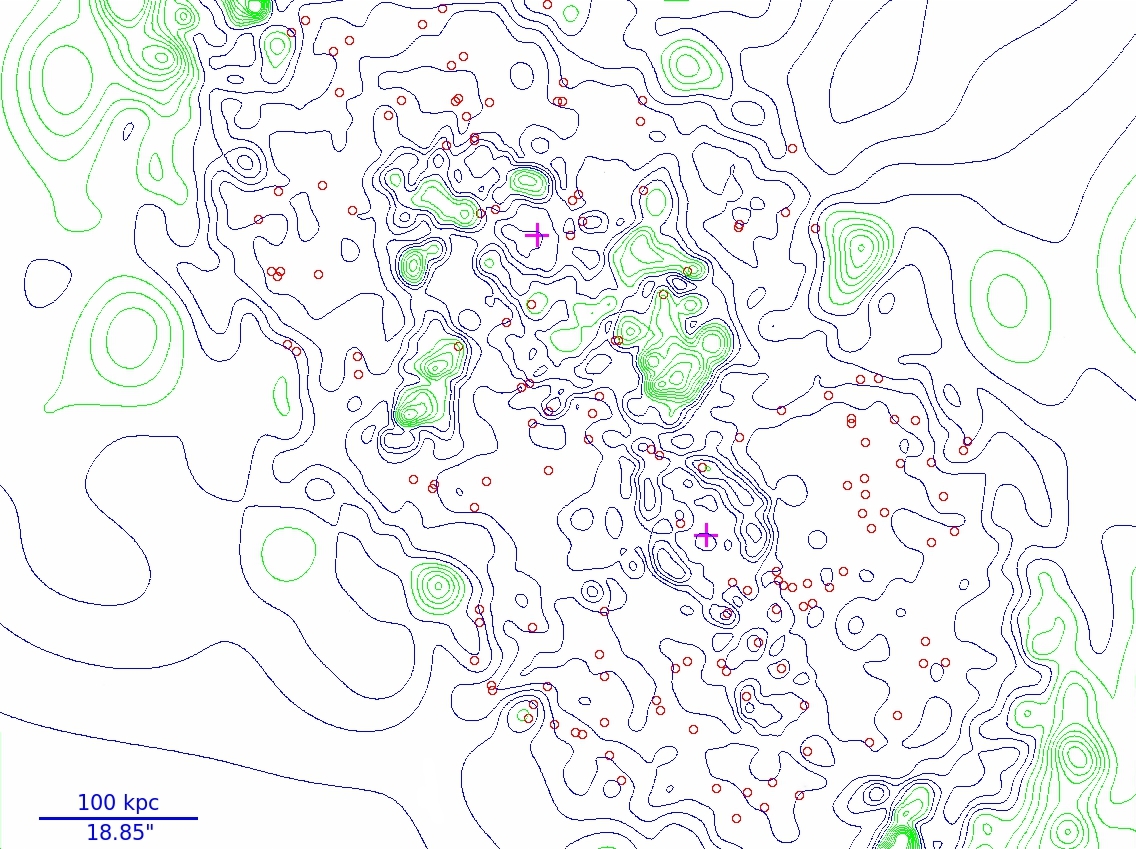}
  \caption{Contours of fractional uncertainty in mass density of MACSJ0416. The contours are separated into two groups distinguishable by their colors. The six blue contour lines range from 3.3\% to 19.8\% in fractional mass density uncertainty, and the seven green lines range from 23.1\% to 49.5\%. Images are highlighted by red circles and the two BCGs are marked as magenta crosses. \label{fig:fracmap}}
\end{figure*}

One way to determine if light traces mass is to analyze the projected mass contours, shown in Fig.~\ref{fig:avgmap}, and their relation to the galaxies. 

%\vskip-0.7in

 It is apparent that the overall mass distribution of the cluster, including its elongation is well reproduced. \grale~finds two prominent cluster-wide mass clumps; these correspond to the two parametric dark matter components used in most of the models of MACSJ0416 described in Section~\ref{sec:lit}. Visual inspection shows that close to the center of the cluster, where the map is best constrained the mass density contours around galaxies encircle the galaxies indicating that \grale~places mass concentration at those locations, even though galaxies are not part of the input. Towards the edges of the cluster, where the mass is least constrained, some contour lines go through galaxies, indicating that these galaxies are not `detected' by \grale.

One of our goals is to determine how well mass follows light in the immediate vicinity of galaxies. These latter galaxies that \grale~does not detect cannot be used for the purpose, but the former ones can. There are four such galaxies along the central ridge line of MACSJ0416; they are marked in Fig.~\ref{fig:avgmap} and have magenta labels G1-G4. These regions also correspond to low fractional uncertainties, $\simlt 10\%$, in reconstructed mass.

%The mass contour is constrained the greatest near the center of the galaxy cluster where most of the images lie. In this region several mass peaks are slightly offset from the closest galaxy positions.

A closer inspection of the mass contours around these four galaxies reveals that the peaks of the mass distribution are displaced from the closest galaxy cluster member. To find out if the displacement is statistically significant we can look at the scatter of the positions of the local density peaks in the 30 individual mass maps. The green dots in Fig.~\ref{fig:avgmap} mark the highest single mass pixel in each of the 30 reconstructions, in circular regions of radius 8\arcsecond~around the four galaxies. Even though the galaxy positions are not centered on the corresponding cloud of 30 mass peaks, they are also not significantly offset from that distribution. Table \ref{tab:offsets} shows the distance in arcseconds between galaxy and average position of the 30 mass peaks. The significance can be defined as the ratio between the offset and root-mean-square dispersion. The last column of Table \ref{tab:offsets} shows none of the galaxies to be significantly offset from the mass peaks. We conclude that there is no compelling evidence that the mass in the immediate vicinity of the central galaxies does not trace the light. This is in contrast to the case of galaxy N1 in ACO 3827 \citep{ws11,moh14,mas15}, where the displacement between the galaxy and the nearest mass peak is $0.89^{+0.26}_{-0.27}$ \arcsecond, or $1.62^{+0.47}_{-0.49}$ kpc.

\begin{table}
  \caption{Offsets between galaxies G1-G4 and 30 mass models peaks in arcseconds. The third column is the root-mean-square dispersion between the galaxies and mass peaks. The last column is the significance, ratio between the distance and root-mean-square.}
  \label{tab:offsets} 
  \begin{tabular}
    { c | c | c | c}
    \hline
    Galaxy & Dist. (\arcsecond) & RMS (\arcsecond) & $\frac{\rm Dist.}{\rm RMS}$ \\ \hline
    
    G1 & 2.77 & 3.62 & 0.77 \\
    G2 & 0.94 & 2.16 & 0.43 \\
    G3 & 5.94 & 3.62 & 1.64 \\
    G4 & 2.20 & 2.35 & 0.94 \\ \hline
    
  \end{tabular}
\end{table}

The width of the distribution of the 30 local mass peaks from individual reconstructions, i.e. our uncertainty, is $\sim\!5$\arcsecond, and represents the smallest offset we could have detected, if these were present. Given that the galaxy-mass offset in ACO 3827, $\sim\!0.9$\arcsecond, yields dark matter self-interaction cross section that is approximately the same as the upper limit from other studies \citep{clo06,ran08,kah14,kah15}, means that the level of uncertainty in MACSJ0416 will not allow a more stringent upper limit on dark matter self-interaction cross section. Although the total number of images in MACSJ0416 is high, $\approx$150, it appears that their distribution, for example the proximity to the four galaxies, or the accuracy of the source redshifts, are not adequate to constrain the offsets at a level comparable to those seen in ACO 3827. The role of image (or image knot) number density in accurately constraining mass maps was discussed in \cite{lie08}. The authors showed that the monopole degeneracy, a way to redistribute mass between images by adding circularly symmetric density distributions with zero total mass, does not change image positions, and is largely responsible for uncertainty in the mass determination. The prevalence of monopole degeneracy will depend on whether the distribution of images allows adding such circular regions. It is possible that this is easier to do near the galaxies G1-G4 in MACSJ0416 than in ACO 3827.  It is possible that in other HFF clusters the configuration of the images is more fortuitous for the detection of possible offsets.

%\begin{figure}
%  \includegraphics[width=\columnwidth]{peakpoints.jpeg}
%  \caption{Local mass peaks around the 4 central ridge galaxies, in 30 individual mass reconstructions. These galaxies are labelled G1-G4, with G1 appearing the brightest.
% \label{fig:cscatter}}
%\end{figure}

\subsection{Mass-Galaxy Correlation Function: \grale}

A further measure of how well mass follows light is provided by the correlation function, $\xi(\theta)$ between galaxies and the average reconstructed mass map. The projected galaxy-mass correlation function describes how galaxies and mass are clustered, as a function of separation, $\theta$,  on the sky. It is defined through conditional probability, $dP$ of finding a galaxy in a volume $dV$, a distance $\theta$ away from another galaxy, $dP=n(1+\xi)dV$, where $n$ is the average number density of galaxies. It typically decreases with separation, after attaining the largest amplitude at zero. We use the estimator $\xi(\theta)=\frac{D_mD_g}{\langle D_mR_g\rangle}-1$,  where $D_mD_g$ represents the number of mass pixel--galaxy pairs, and $\langle D_mR_g\rangle$ is the average number of pairs of 100 trials, where the positions of the galaxies have been randomized.

\begin{figure}
  \includegraphics[width=\columnwidth]{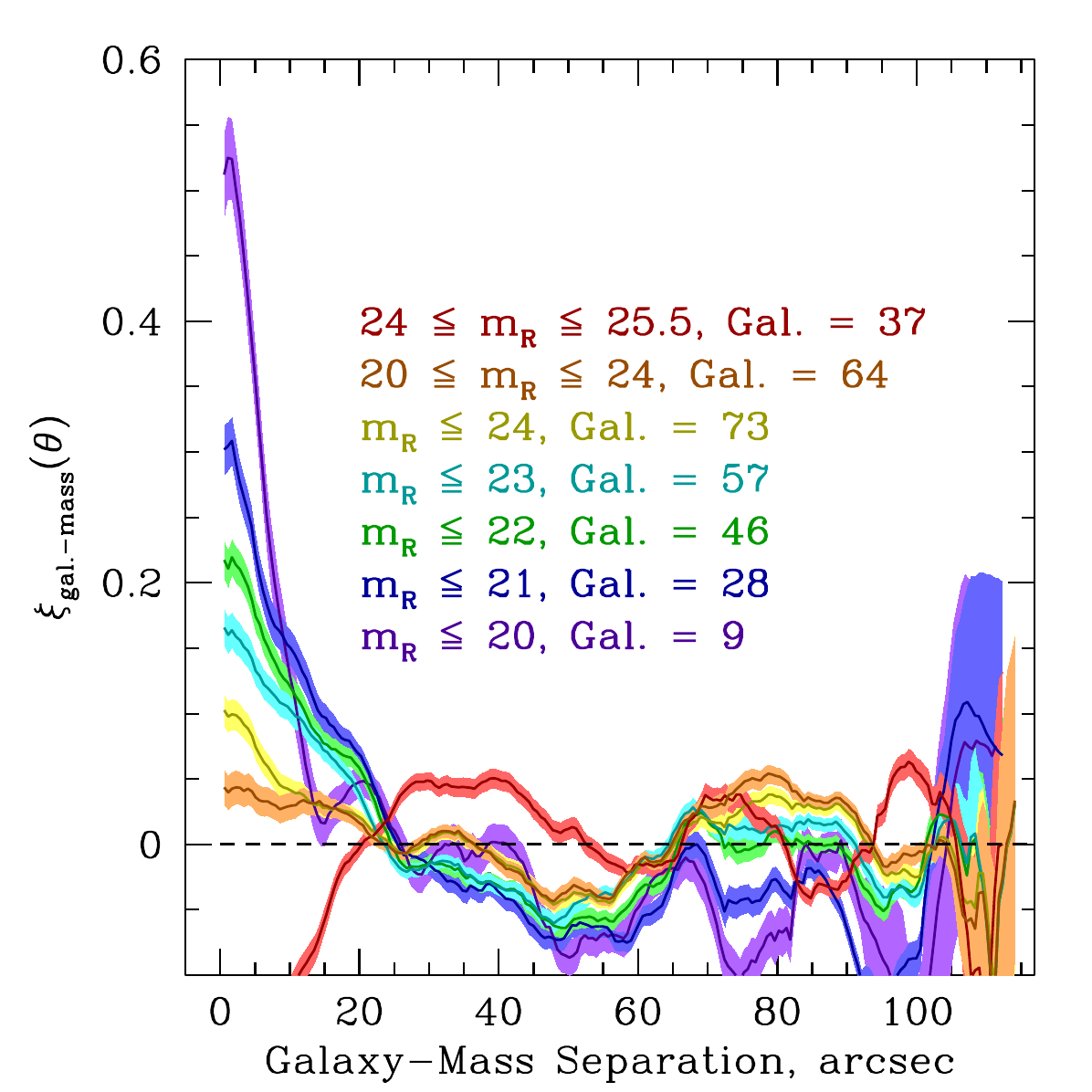}
  \caption{Normalized galaxy-mass cross-correlation function for Subaru R-band at multiple magnitude cuts. The mass map is the average of 30 individual \grale~reconstructions. Rms dispersion between the 100 realizations of $D_m R_g$ is approximately 0.025 for all magnitude ranges. Shaded regions indicate $1\sigma$ error bars.} \label{fig:corr}
\end{figure}

The galaxy-mass correlation function between the galaxies in the Subaru R-band catalog in MACSJ0416 and the average mass map is shown in Fig.~\ref{fig:corr}. It was computed using a bin size of $0.52$\arcsecond within a region of area $5903.52$ squared arcseconds and enclosing galaxies and images along the cluster's line of elongation. This is the line extending through the two brightest cluster galaxies of the cluster. We chose Subaru galaxies over HST galaxies because Subaru catalog contains magnitudes for all HST galaxies, including those that are too bright for HST. Furthermore, Subaru filters we use are closely matched by HST filters; specifically, galaxy magnitudes in Subaru Z (R) band are tightly correlated with HST F814W (F606W) magnitudes. Since Subaru is a ground-based telescope, the light profiles of the brightest galaxies are extended, and may block fainter galaxies. To reduce this bias, we masked the 10 brightest galaxies in the Subaru R-band with a circle of radius 8" before calculating the correlation functions. Fig.~\ref{fig:corr} shows $\xi(\theta)$ for several galaxy magnitude cuts; the number of galaxies in each cut is indicated in the figure. 

If galaxies traced the underlying mass distribution according to the standard biasing scheme one would expect that the brightest galaxies would show the strongest clustering with the mass, and as the galaxy magnitude limit is pushed towards lower fluxes, the amplitude of the correlation function would decrease due to two reasons: fainter galaxies contain less mass, and are thus biased towards less mass. This is indeed what is seen. Because \grale~ input does not include any information about the galaxies, the trend of brighter galaxies being more biased towards mass is an important confirmation of the standard biasing scenario.

Fig.~\ref{fig:corr} also shows that at very small separations, $\simlt 2$\arcsecond, the correlation function decreases, instead of increasing as separation approaches zero. Taken at face value, this would be indicative of an offset between dark and light matter in cluster galaxies. However, the uncertainties in $\xi$, which are approximately 0.025, are too large to make that conclusion. This behaviour at small separations and its marginal statistical significance was already seen in Section~\ref{sec:peaks}.

\begin{figure}
  \includegraphics[width=\columnwidth]{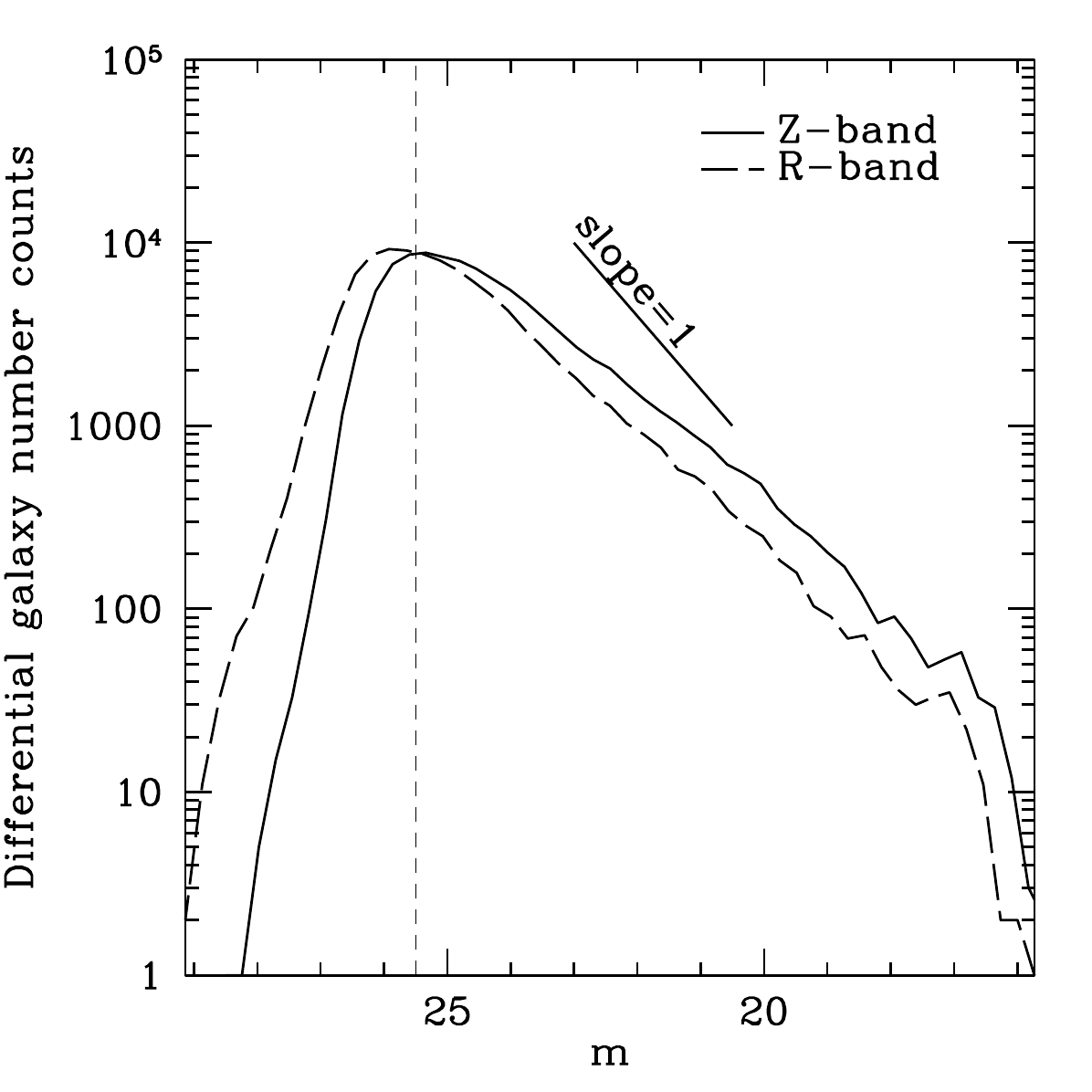}
  \caption{Differential number counts of galaxies over the entire Subaru field, not just the galaxies behind the cluster. The vertical dashed line marks the magnitude below which counts are incomplete; we did not include these galaxies in any of the analysis. The marked line, $d\log(n[m])/d\log(f)=1$, is provided for comparison to the two bands.}
  \label{fig:numcount}
\end{figure}

While it is safe to assume that bright galaxies are mostly cluster members, the same cannot be said of fainter ones. One way to find the apparent magnitude below which most galaxies are background to the cluster is to use lensing magnification bias. Behind a galaxy cluster lensed galaxies are made brighter than unlensed ones at the same redshifts and the area behind the lens is simultaneously stretched by the same magnification factor.  To predict the net effect of this bias we can look at the differential galaxy counts as a function of apparent magnitude. Fig.~\ref{fig:numcount} shows these counts based on the entire Subaru field; using just the galaxies in the direction of the cluster would produce very noisy counts. Magnification bias decreases galaxy counts at magnitudes where the slope of the counts is shallow, and increases them at magnitudes where the slope is steep.  When the slope is equal to one, $d\log(n[m])/d\log(f)=1$, there is no magnification bias because the flux magnification and area dilution cancel each other out. In principle, one should look at the unlensed counts slope for this purpose, which are unobservable, but in the case of relatively shallow counts, the lensed observed counts provide a reasonable approximation.

Because at magnitudes where $d\log(n[m])/d\log(f)<1$ area dilution wins over flux magnification, magnification bias predicts anti-correlation between cluster mass and background galaxies. According to Fig.~\ref{fig:numcount}, $d\log(n[m])/d\log(f)$ becomes shallower than 1 below $m\approx 20$, for both R and Z magnitudes. We select two magnitude cuts, $20\!\le\!m\!\le\!24$ and $24\!\le\!m\!\le\!25.5$; we did not use galaxies fainter than 25.5 because beyond this magnitude counts suffer significantly from incompleteness.

\begin{figure*}
%  \centering
%\vskip-0.75in
%\hskip-0.75in
  \includegraphics[width=\textwidth]{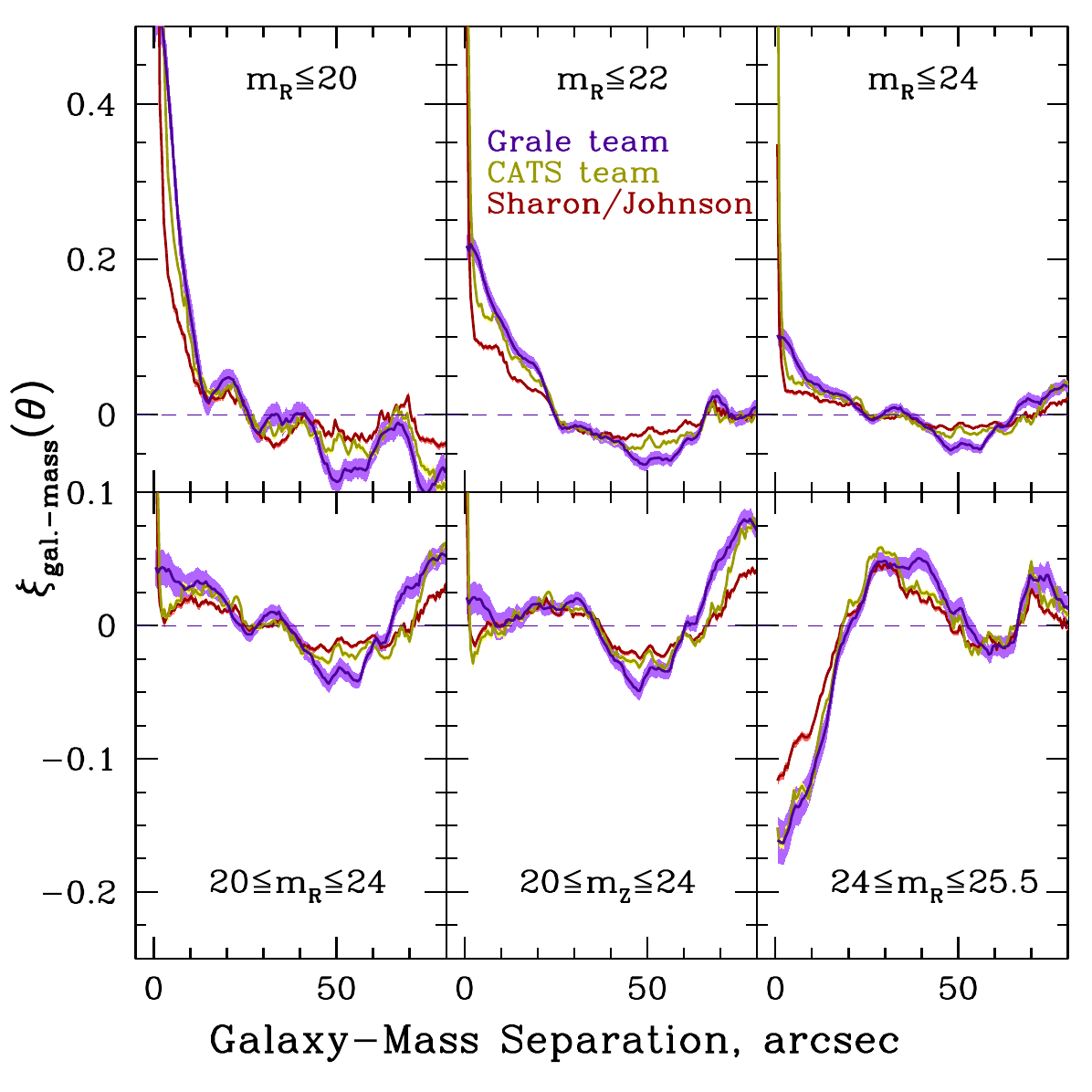}
%\vskip-1.75in
  \caption{Galaxy-mass cross-correlation functions for five Subaru R-band and one Z-band magnitude cut. Three mass maps are used: from the \grale~team (violet), CATS team (gold), and Sharon/Johnson (red). Shaded regions indicate $1\sigma$ error bars.}
  \label{fig:6panels}
\end{figure*}

For the $20\!\le\!m\!\le\!24$ cut we expect modest anti-correlation for the R-band, and somewhat larger anti-correlation for the Z-band, because Z-band counts are shallower than those in R-band. The same masking procedure was applied to the Z-band for computing the correlation function. The computed correlation functions for both bands are shown in violet in the lower middle left and right panels of Fig.~\ref{fig:6panels}. Indeed, the Z-band galaxies show less correlation than the R-band galaxies. However, neither the R-band selected galaxies nor the Z-band selected galaxies show an anti-correlation between \grale~cluster mass and galaxies. This might indicate that some of the galaxies in this magnitude range are cluster members, and so the total correlation function signal consists of a superposition of mass clustering with the cluster member galaxies, and magnification bias resulting from the cluster mass lensing background galaxies.

At fainter galaxy magnitudes, $24\!\le\!m\!\le\!25.5$, we expect that most, if not all galaxies are background to the cluster, and so magnification bias will be the only effect. Consistent with this expectation we find a strong anti-correlation between cluster mass and cluster galaxies for both Subaru magnitude bands (we show only R band results here). Thus, different magnitude cuts show features that imply that magnification bias exists in MACSJ0416, and so many of the fainter galaxies are significantly background to the cluster. Magnification bias is sometimes used to aid mass reconstruction in clusters \citep{ume08,ume15}.

We note that the (anti-)correlations extend to about $20$\arcsecond, or 100 kpc at the redshift of the cluster (Fig.~\ref{fig:corr}), which is roughly the typical separation between bright, $m\simlt 20$ galaxies, and considerably smaller than the size of the cluster, $120$\arcsecond$\times 50$\arcsecond.

\subsection{Mass-Galaxy Correlation Function: \grale~vs. \lenstool}

A number of metrics can be used to compare mass reconstructions from different lens inversion methods. One can look at the differences, or fractional differences between two $\kappa$ maps on the $x,y$ plane, however, summarising that information in a concise way is difficult. One can also look at the circularly averaged radial density profiles of the different maps, however, this entails a lot of averaging, which hides many potentially interesting differences. We use the correlation function, which is similar to circular averaging, but is done around each galaxy, instead of just the center of the cluster. It is a good compromise between too much and too little detail.

We compare \grale~results to those of two groups that use different implementations of \lenstool, CATS team and Sharon/Johnson team, which are presented in \cite{jau15} and \cite{joh14}, respectively, and are available for download on the HST MAST website. In this work, we limit our comparison to reconstructions based on \lenstool~only, because \grale~and \lenstool~are at the opposite ends of the spectrum of assumptions going into lens inversion methods. We leave a broader comparison to a later work. The CATS reconstruction is based on the HFF strong and weak lensing data, and Sharon/Johnson reconstruction uses pre-HFF strong lensing data. Both assume flat $\Lambda$CDM cosmology, with $\Omega_m=0.3$, and $h=0.7$. This is different from our assumed cosmology, but the difference results in negligible deviations of length scales, $\sim$ 0.5$\%$.

Figure~\ref{fig:6panels} plots mass-galaxy correlation function for six different galaxy magnitude cuts. All but the bottom middle panel use Subaru R-band to select galaxies; the bottom middle panel uses Z-band. The magnitude cuts in the six panels are $m_R\leq 20$, $m_R\leq 22$, $m_R\leq 24$, $20\leq m_R\leq24$,  $20\leq m_Z\leq24$, and  $24\leq m_R\leq25.5$.  A cursory look at these reveals that all three teams recover very similar clustering properties. The fact that parametric and free-form methods yield very similar results is encouraging, and leads to two conclusions, (a) strong priors about mass following light are not required to recover mass distribution in clusters with very many lensed images, (b) light follows mass quite well in this merging cluster.

Though \grale~and the two implementations of \lenstool~agree overall quite well, there are notable differences between the three.

CATS and Sharon/Johnson correlation functions have a sharp spike near zero separations. This tight correlation is the result of \lenstool~placing mass at the locations of the visible galaxies. The central spike in \lenstool~models can be also seen in the projected mass power spectrum of the cluster studied by \cite{moh16}: see the left panel of their Fig.~9, where the fluctuation power at large $k$ does not fall as fast as \grale's. \grale, which is blind to galaxies, still detects galaxies as gauged by this metric, but does not associate as much mass with them due to a combination of two reasons, (i) the positions of lensed images do not require it, and (ii) \grale~does not have sufficient spatial resolution.

Aside from the central spike, the correlation functions of the two implementations of \lenstool, CATS and Sharon/Johnson, disagree with each other at the same level as they disagree with \grale. In fact, in the range $\theta=2$\arcsecond$-20$\arcsecond~CATS and \grale~are closer to each other than CATS and Sharon/Johnson. This behaviour is probably the result of the data sets used: \grale~and CATS use HFF, while Sharon/Johnson use pre-HFF. 

Given that CATS and \grale~teams use the same strong lensing data sets, the difference between these two should be entirely due to the differences in the reconstruction method. A closer examination shows that at separations $2$\arcsecond$-10$\arcsecond~and magnitude cuts above $m_R=24$, CATS team correlation functions' amplitude is below that of \grale---one can say it shows a dip---and below a smooth extrapolation of $\xi(\theta)$ from larger separations. On these scales \grale's $\xi(\theta)$ looks smooth and rises monotonically in that region. In other words, compared to \grale, CATS \lenstool~mass maps show a steeper decline in correlation function with increasing separation. This is likely a consequence of how \lenstool~builds its mass maps as a superposition of large, smooth dark matter component(s) and mass spikes due to galaxies. There is no intermediate mass component that could bridge the gap and introduce more mass on these length scales around galaxies.

\grale's $\xi(\theta)$ gentler decline with $\theta$ on scales $2$\arcsecond$-10$\arcsecond~or $10-50$ kpc may suggest that there is more galaxy-mass correlation on these scales, i.e. that galaxies, and especially those in the magnitude range $22\simlt m_R\simlt 24$, are more biased traces of mass in clusters than \lenstool~assumes.  In this context it is interesting to recall a recent discovery of nearly a thousand low surface brightness galaxies with $1$~kpc~$\simlt R_e\simlt 3$~kpc in the center of Coma cluster \citep{kod15}. Since these galaxies are old and evolved, they likely existed in clusters at  higher redshifts, comparable to those of HFF clusters. Though faint and likely low mass, they still outnumber cluster members. These are not part of \lenstool~input, and if they cluster with brighter galaxies, they could contribute to higher correlation amplitude on scales of tens of kpc. 

In general, the somewhat different galaxy-mass clustering amplitudes found by the three groups imply distinct models of how mass is distributed in clusters, and have different implications for the cluster structure and evolution. For example, a larger galaxy-mass clustering amplitude, such as obtained by \lenstool~vs \grale, may imply that galaxies form at much higher density peaks of the matter distribution, or that only the most compact galaxies can survive in cluster centers. Hydrodynamic numerical simulations could help to further determine the implications of the degree of biasing within clusters.

\section{Conclusions}
\label{sec:con}

We carried out a lens inversion of MACSJ0416 using \grale, a free-form genetic algorithm based method. The only inputs were 149 lensed images, identified by \cite{jau14} based on HFF data. First, we summarize our results on the comparison between \grale~and the two \lenstool~mass reconstructions of this massive, merging $z\sim 0.4$ cluster. Because the true mass distribution in galaxy clusters is unknowable, it is important to critically compare the commonly used estimators. Our conclusions based on these two very different methodologies are:

\begin{itemize}

\item Though the mass maps of \grale~and \lenstool~reconstructions look somewhat different in detail, \grale~and \lenstool~(specifically, CATS and Sharon/Johnson teams) recover very similar statistics of the mass-galaxy correlation.

\item The most striking, but expected difference is that \lenstool's galaxy-mass correlations show a pronounced spike near zero separation. This is because \lenstool~places a lot of mass at the locations of galaxies as part of its input, whereas \grale~does not. 

\item Another notable difference is that on scales of $2$\arcsecond$-10$\arcsecond, or tens of kpc, \grale's correlation function falls less steeply than \lenstool's. It is possible that the lensing constraints are compatible with both the steeper and the shallower decline. However, it is also possible that the mass distribution is more extended on these scales around galaxies, as \grale~suggests, due to the presence of many hundreds of low surface brightness galaxies of the kind recently detected in Coma \citep{kod15}.

\end{itemize}

The fact that two very different methodologies---\grale~and \lenstool---give similar results leads us to conclude that when lensed image number is around 100 or more, the images alone are sufficient to recover the mass distribution in clusters very well. Strong priors on galaxies are not needed.

Our conclusions regarding the mass vs. light distribution in MACSJ0416 are:

\begin{itemize}

\item Even with the high number of images present in MACSJ0416, no significant mass-light offsets are found between the four central galaxies and the nearest mass peaks, in contrast to ACO 3827. Our uncertainties are $\sim 5$\arcsecond, larger than $0.9$\arcsecond~offset observed in ACO 3827.  It is possible that other HFF clusters have a more fortuitous image configuration and hence smaller uncertainties around the bright cluster galaxies.

\item Overall, on scales larger than a few arcseconds, light traces mass, as reconstructed by \grale, in the merging cluster MACSJ0416 quite well, as measured by the galaxy-mass correlation function. This is the only analysis of this merging cluster that does not use any information about the visible light, hence the conclusion that light follows mass is not trivial.

\item The faintest galaxies in the direction of MACSJ0416 are anti-correlated with the cluster mass, implying the presence of the lensing magnification bias

\end{itemize}

\section*{Acknowledgements}

KS and LLRW acknowledge the computational resources and support of the Minnesota Supercomputing Institute. JL acknowledges the use of the computational resources and services provided by the VSC (Flemish Supercomputer Center), funded by the Hercules Foundation and the Flemish Government - department EWI. LLRW would like to thank Nina A. Rodrigues for bringing the work of Jin Koda and collaborators to her attention. IM is supported by Fermi Research Alliance, LLC under Contract No. De-AC02-07CH11359 with the United States Department of Energy.

%%%%%%%%%%%%%%%%%%%%%%%%%%%%%%%%%%%%%%%%%%%%%%%%%%


% Don't change these lines
\bsp	% typesetting comment
\label{lastpage}

\begin{thebibliography}{99}

\bibitem[Albada(1961)]{alb61}
Albada, G.B. van, 1961, AJ, 66, 590
% Evolution of Clusters of Galaxies under Gravitational Forces

\bibitem[Blumenthal et al.(1984)]{blu84}
Blumenthal, G. R., Faber, S. M., Primack, J. R., and Rees, M. J. 1984. Nature, 311, 517
% Formation of galaxies and large-scale structure with cold dark matter. 

\bibitem[Bullock et al.(2012)]{bul12}
Bullock, J. et al. 2012, ``Hubble Deep Fields Initiative 2012 – Science Working Group Report''

\bibitem[Clowe et al.(2006)]{clo06}
Clowe, D., Bradac, M., Gonzalez, A. H., Markevitch, M., Randall, S. W., Jones, C., Zaritsky, D
2006, ApJ, 648, L109
%A Direct Empirical Proof of the Existence of Dark Matter

\bibitem[Coe et al.(2012)]{coe12}
Coe, D., et al. 2012, ApJ, 757, 22
% CLASH: Precise New Constraints on the Mass Profile of the Galaxy Cluster A2261

\bibitem[Diego et al.(2015)]{die15}
Diego J. M., Broadhurst T., Molnar S. M., Lam D., Lim J., 2015, MNRAS,
447, 3130
% Free-form lensing implications for the collision of dark matter and gas in the frontier fields cluster MACS J0416.1-2403

\bibitem[Fort \& Mellier(1994)]{fm94}
Fort, B. \& Mellier, Y. Astron. \& Astrophys. Rev., 5, 239
% Arc(let)s in clusters of galaxies

\bibitem[Grillo et al.(2015)]{gri15}
C. Grillo et al. 2015, ApJ, 800, 38
% CLASH-VLT: Insights on the mass substructures in the Frontier Fields Cluster MACS J0416.1-2403 through accurate strong lens modeling

\bibitem[Jauzac et al.(2014)]{jau14}
Jauzac M. et al., 2014, MNRAS, 443, 1549
% Hubble Frontier Fields: a high-precision strong-lensing analysis of galaxy cluster MACSJ0416.1-2403 using ∼200 multiple images 

\bibitem[Jauzac et al.(2015)]{jau15}
Jauzac M. et al., 2015, MNRAS, 446, 4132
% Hubble Frontier Fields: the geometry and dynamics of the massive galaxy cluster merger MACSJ0416.1-2403

\bibitem[Johnson et al.(2014)]{joh14}
{Johnson}, T.~L., {Sharon}, K., {Bayliss}, M.~B., {Gladders}, M.~D., {Coe}, D. \& {Ebeling}, H. 2014, ApJ, 797, 48
% Lens Models and Magnification Maps of the Six Hubble Frontier Fields Clusters

\bibitem[Jullo et al.(2007)]{jul07}
Jullo E., Kneib J.-P., Limousin M., Elıasdottir A., Marshall P. J., Verdugo T., 2007, New Journal of Physics, 9, 447

\bibitem[Kahlhoefer et al.(2014)]{kah14}
Kahlhoefer, F., Schmidt-Hoberg, K., Frandsen, M., Sarkar, S., 2014, MNRAS, 437 2865
% Colliding clusters and dark matter self-interactions

\bibitem[Kahlhoefer et al.(2015)]{kah15}
Kahlhoefer F., Schmidt-Hoberg K., Kummer J., Sarkar S., 2015, MNRAS,
452, L54
% On the interpretation of dark matter self-interactions in Abell 3827

\bibitem[Koda et al.(2015)]{kod15}
Koda J., Masafumi Y., Hitomi Y., Yutaka K., 2015, ApJ, 807, L2
% Approximately a Thousand Ultra-diffuse Galaxies in the Coma Cluster

\bibitem[Liesenborgs et al.(2006)]{lie06}
Liesenborgs, J., de Rijcke, S. \& Dejonghe, H. 2006, MNRAS, 367, 1209
%{\it A genetic algorithm for the non-parametric inversion of strong lensing systems}

\bibitem[Liesenborgs et al.(2007)]{lie07}  % 0707.2538
Liesenborgs, J., de Rijcke, S., Dejonghe, H. \& Bekaert, P. 2007, MNRAS, 380, 1729
%{\it Non-parametric inversion of gravitational lensing systems with few images using a multi-objective genetic algorithm}

\bibitem[Liesenborgs et al.(2008)]{lie08}
Liesenborgs, J., de Rijcke, S., Dejonghe, H., Bekaert, P. 2008, MNRAS, 389, 415
%{\it Non-parametric strong lens inversion of Cl 0024+1654: illustrating the monopole degeneracy}

\bibitem[Liesenborgs et al.(2009)]{lie09}
Liesenborgs, J., de Rijcke, S., Dejonghe, H., Bekaert, P. 2009,MNRAS, 397, 341
%{\it Non-parametric strong lens inversion of SDSS J1004+4112}

%\bibitem[Liesenborgs \& De Rijcke(2012)]{lie12}  % arXiv:1207.4692 
%Liesenborgs, L. \& De Rijcke, S. 2012, MNRAS, 425, 1772
%{\it Lensing degeneracies and mass substructure}

\bibitem[Kneib \& Natarajan(2011)]{kn11}
{Kneib}, J.-P. \& {Natarajan}, P. 2011, Astron. \& Astrophy. Rev., 19, 47
% Cluster lenses

\bibitem[Mann \& Ebeling(2012)]{man12}
Mann, A. W., \& Ebeling, H. 2012, MNRAS, 420, 2120

\bibitem[Massey et al.(2015)]{mas15}
Massey, R., Williams, L.L.R., Smit, R., Swinbank, M., Kitching, T. Harvey, D. Israel, H.,
Jauzac, M., Clowe, D., Edge, A., Hilton, M., Jullo, E., Leonard, A., Liesenborgs, J., Merten, J.,
Mohammed, I., Nagai, D., Richard, J., Robertson, A., Saha, P., Santana, R., Stott, J., Tittley, E.  2015, MNRAS, 449, 3393
%{\it The behaviour of dark matter associated with 4 bright cluster galaxies in the 10 kpc core of A3827}

\bibitem[Merten et al.(2011)]{mer11}
Merten, J. et al. 2011, MNRAS, 417, 333
% Creation of cosmic structure in the complex galaxy cluster merger Abell 2744

\bibitem[Mohammed et al.(2014)]{moh14}
Mohammed, I., Liesenborgs, J., Saha, P., Williams, L.L.R. 2014, MNRAS, 439, 2651
%{\it Mass-Galaxy offsets in A3827, 2218 \& 1689: intrinsic properties or line-of-sight substructures?}

%\bibitem[Mohammed et al.(2015a)]{moh15a}
%Mohammed, I., Saha, P.,  Liesenborgs, J. 2015, PASJ, 67, 21
%{\it Lensing time delays as a substructure constraint: a case study with the cluster SDSS J1004+4112}

\bibitem[Mohammed et al.(2016)]{moh16}
Mohammed I., Saha P., Williams L. L. R., Liesenborgs J., Sebesta K., 2016, MNRAS, 459, 1698
%{\it Quantifying sub-structures in Hubble Frontier Field clusters: comparison with $\Lambda$CDM simulations}

\bibitem[Ogrean et al.(2015)]{ogr15}
Ogrean G. et al., 2015, ApJ, 812, 153
% Frontier Fields Clusters: Chandra and JVLA View of the Pre-Merging Cluster MACS J0416.1-2403

\bibitem[Peebles(1970)]{pee70}
Peebles, P.J.E. 1970, AJ, 75, 13
% Structure of the Coma Cluster of Galaxies

\bibitem[Peebles(1983)]{pee83}
Peebles, P.J.E. 1983, ApJ, 274, 1
% The sequence of cosmology and the nature of primeval departures from homogeneity

\bibitem[Randall et al.(2008)]{ran08}
Randall, S. W., Markevitch, M., Clowe, D., Gonzalez, A. H.; Bradac, M. 2008, ApJ, 679, 1173  %0704.0261
%Constraints on the Self-Interaction Cross-Section of DM from Numerical Sim of the Merging Galaxy Cluster 1E 0657-5 

\bibitem[Richard et al.(2014)]{ric14}
Richard, J. et al. 2014, MNRAS, 444, 268
% Mass and magnification maps for the Hubble Space Telescope Frontier Fields clusters: implications for high-redshift studies

%\bibitem[Saha(2000)]{sah00}
%Saha, P. 2000, AJ, 120, 1654
%{\it Lensing Degeneracies Revisited}

\bibitem[Umetsu \& Broadhurst(2008)]{ume08} 
Umetsu, K. \& Broadhurst, T. 2008, ApJ, 684, 177
% Combining Lens Distortion and Depletion to Map the Mass Distribution of A1689

\bibitem[Umetsu et al.(2015)]{ume15}
Umetsu K., Zitrin A., Gruen D., Merten J., Donahue M., Postman M., 2015,
ApJ, 821, 116
% CLASH: Joint Analysis of Strong-Lensing, Weak-Lensing Shear and Magnification Data for 20 Galaxy Clusters

\bibitem[Williams \& Saha(2011)]{ws11}
 Williams, L.L.R., Saha, P. 2011, MNRAS, 415, 448
% Light/mass offsets in the lensing cluster Abell 3827: evidence for collisional dark matter?

\bibitem[Zitrin et al.(2010)]{zit10}
Zitrin, A. et al. 2010, MNRAS, 408, 1916
% {\it Full lensing analysis of Abell 1703: comparison of independent lens-modelling techniques.}

\bibitem[Zitrin et al.(2013)]{zit13}
Zitrin, A. et al. 2013, ApJ, 762, 30
% CLASH: The Enhanced Lensing Efficiency of the Highly Elongated Merging Cluster MACS J0416.1-2403}

%\bibitem[Zitrin et al.(2009)]{zit09}   % image identification proceedure  % arXiv:0902.3971
%Zitrin, A., Broadhurst, T., Umetsu, K., Coe, D., Benitez, N., Ascaso, B., Bradley, L., Ford, H., et al. 2009, 
%MNRAS, 396, 1985
%{\it New multiply-lensed galaxies identified in ACS/NIC3 observations of Cl0024+1654 using an improved mass model}



\end{thebibliography}
\end{document}